\documentclass[letter]{IEEEtran}
\usepackage{amsfonts}
\usepackage{graphicx}
\usepackage{color}
\usepackage{amsmath,amsfonts,amssymb,amsthm,epsfig,epstopdf,url,array}
\usepackage{url,textcomp}
\usepackage{authblk}
\usepackage{cite}
\usepackage{verbatim}
\usepackage{subfigure}
\usepackage{caption}
\newcommand{\bs}{\boldsymbol}
\newtheorem{theorem}{Theorem}

\def\BibTeX{{\rm B\kern-.05em{\sc i\kern-.025em b}\kern-.08em
    T\kern-.1667em\lower.7ex\hbox{E}\kern-.125emX}}
\makeatletter

\newcommand{\Rmnum}[1]{\expandafter\@slowromancap\romannumeral #1@}
\makeatother
\begin{document}
\title{Outage Performance of Cross-Packet HARQ}
\author{Jiahui~Feng,
        Zheng~Shi,
        Guanghua Yang,
        Nikolaos I. Miridakis,
        Shaodan Ma,
        and Theodoros A. Tsiftsis
\thanks{Manuscript received January 8, 2022; revised April 15, 2022; accepted May 1,2022. This work was supported in part by National Natural Science Foundation of China under Grants 62171200 and 62171201, in part by Guangdong Basic and Applied Basic Research Foundation under Grant 2019A1515012136, in part by Zhuhai Basic and Applied Basic Research Foundation under Grant ZH22017003210050PWC, in part by the Science and Technology Development Fund, Macau SAR (File no. 0036/2019/A1, 0032/2019/AGJ, and File no. SKL-IOTSC(UM)-2021-2023, in part by 2018 Guangzhou Leading Innovation Team Program (China) under Grant 201909010006, and in part by the Research Committee of University of Macau under Grant MYRG2018-00156-FST.
\emph{(Corresponding Author: Zheng Shi.)}}
\thanks{Jiahui~Feng, Zheng Shi, Guanghua Yang, and Theodoros A. Tsiftsis are with the School of Intelligent Systems Science and Engineering, Jinan University, Zhuhai 519070, China (e-mails: jiahui@stu2020.jnu.edu.cn; zhengshi@jnu.edu.cn; ghyang@jnu.edu.cn;  theo\_tsiftsis@jnu.edu.cn).}
\thanks{Nikolaos I. Miridakis is with the school of Intelligent Systems Science and Engineering, Jinan University, Zhuhai 519070, China, and also with the Dept. of Informatics and Computer Engineering, University of West Attica, Aegaleo 12243, Greece (e-mail: nikozm@uniwa.gr).}
\thanks{Shaodan Ma is with the State Key Laboratory of Internet of Things for Smart City, University of Macau, Macau, China (e-mail: shaodanma@um.edu.mo).}
}
\maketitle
\begin{abstract}
As opposed to hybrid automatic repeat request with incremental redundancy (HARQ-IR) that all the resources are occupied to resend the redundant information, cross-packet HARQ (XP-HARQ) allows the introduction of new information into retransmissions to substantially exploit the remaining resources. This letter provides a profound investigation into the outage performance of XP-HARQ. In particular, the exact outage expression of XP-HARQ is derived if the maximum number of transmissions is two, and tight outage bounds are obtained for more than two transmissions. Moreover, the asymptotic outage analysis of XP-HARQ in the high signal-to-noise ratio (SNR) regime is carried out not only to simplify the outage expression, but also to show that full time diversity is achievable by XP-HARQ. The simulation results are eventually presented for verifications.
\end{abstract}
\IEEEpeerreviewmaketitle
\hyphenation{HARQ}
\section{Introduction}\label{sec:int}
\IEEEPARstart As a key feature of the fifth generation (5G), ultra-reliable low-latency communications (URLLC) have widespread applications in industrial automation, self-driving, augmented reality, etc. \cite{8636206,9662195}. However, it is somewhat contradictory to achieve high reliability and low latency simultaneously, which challenges the realization of URLLC. For instance, in order to meet the high requirement of reliability, hybrid automatic repeat request (HARQ) has been widely used in contemporary communication systems. 
More specifically, HARQ can be implemented by using different encoding and decoding strategies, which result in three categories, including Type-I HARQ, HARQ with chase combining (CC-HARQ) and HARQ with incremental redundancy (INR-HARQ) \cite{ahmed2021hybrid}. In essence, HARQ improves the reliability at the expense of large transmission delay. 
Therefore, it is imperative to develop evolved HARQ mechanisms to satisfy more stringent and flexible requirements of reliability and latency.

To shorten communication delay, HARQ and multiple access technologies are integrated to form multi-packet HARQ (MP-HARQ), which allows the delivery of multiple packets in one HARQ round by sharing the same resource block \cite{jabi2015multipacket}. So far, two main resource sharing modes have been proposed for MP-HARQ, including superposition coding (SC) and joint encoding, which are implemented in power domain and code domain, respectively. 
In particular, the idea of power-domain non-orthogonal multiple access (NOMA) was transferred to realize MP-HARQ in \cite{mheich2020performance}, where the SC was applied to multiplex all users' data. 
Furthermore, in \cite{khreis2020multi}, the SC was used to superimpose the redundant information of failed message and new information, which yields multi-layer HARQ. The achievable rate region of multi-layer HARQ was then analyzed in \cite{khreis2020multi} from the information-theoretical perspective. 
A similar MP-HARQ to \cite{khreis2020multi} was reported in \cite{nadeem2021non}, i.e., non-orthogonal HARQ (N-HARQ). To guarantee packet-level latency and in-time delivery, Faisal Nadeem {\emph{et al.}} in \cite{nadeem2021non} examined the block error rate (BLER) and delay of N-HARQ in the finite block length regime. 
Moreover, the joint encoding is another way to implement MP-HARQ from code domain. In \cite{4202098}, Christoph Hausl {\emph{et al.}} proposed a cross-packet HARQ (XP-HARQ), where the failed packet is concatenated with the new packet to construct a long packet for a retransmission. 
In \cite{jabi2017boost}, Mohammed Jabi {\emph{et al.}} took a step further to consider a general XP-HARQ scheme with more than one retransmissions, and proposed a practical implementation method for XP-HARQ. 
Besides, the outdated channel state information was exploited to maximize throughput of XP-HARQ via rate adaptation in \cite{jabi2018amc}.  
In \cite{liang2018efficient}, polar codes and backtrack-freezing decoding were leveraged to realize XP-HARQ, in which the simulation results indicated that XP-HARQ attains a significant gain over HARQ-IR with less transmission times.
However, the prior available results with regard to the performance of XP-HARQ are based on either simulation or approximation, and lack of meaningful insights. This is due to the significant challenge of tackling the involved expression of the achievable rate region. 

Motivated by the above issue, this letter first derives the exact expression of the outage probability of XP-HARQ for a maximum of two transmissions, with which the asymptotic outage probability at high signal-to-noise ratio (SNR) is obtained. Furthermore, since it is virtually impossible to derive an exact outage expression for XP-HARQ with more than two transmissions, we resort to upper and lower bounds of the outage probability for approximation. Similarly, the asymptotic outage expression is obtained to reveal meaningful insights, such as diversity order. 

The rest of this letter is organized as follows. Section \ref{sec:sys} presents the system model, based on which Section \ref{sec:out} conducts the outage analysis. The theoretical analysis is then validated in Section \ref{NUM_RES}. Finally, Section \ref{sec:con} concludes this letter.

\section{System Model}\label{sec:sys}

Before proceeding to the outage analysis, the system model is presented in this section, including the transmission mechanism of XP-HARQ and the formulation of outage probability.
\subsection{XP-HARQ}
In the first HARQ round, the message ${\rm m}_1$ is first encoded and then modulated as ${\bf x}_1$ with a initial transmission rate $R_1$. 
Thus, the received signal ${\bf y}_1$ is written as ${{\bf y}_1} = \sqrt{P_1}{h_1}{{\bf x}_1} + {{\bf z}_1}$,
where ${P_1}$ is the average transmitted power, ${{\bf z}_1}$ is the complex additive Gaussian white noise (AWGN) with zero mean and variance of ${\sigma ^2}$. ${h_1}$ is the Rayleigh channel coefficient with normalized average power, i.e., ${\rm E}\{{\left| {{h_1}} \right|^2}\} = 1$. If the receiver succeeds in decoding the message ${\rm m}_1$, a positive acknowledgement (ACK) is fed back to the sender and the delivery of the next message ${\rm m}_2$ is launched. On the contrary, the retransmission mechanism is triggered. According to the principle of XP-HARQ \cite{jabi2017adaptive}, 
the previously failed messages are jointly encoded with the subsequent message to avoid the ``waste'' of mutual information. More specifically, during the $k$-th HARQ round, the messages ${\rm m}_1,\cdots,{\rm m}_k$ are combined to form a longer message ${\rm m}_{[k]}$, which is jointly encoded and modulated as a codeword ${{\bf x}_k}$. Thus, the received signal ${{\bf y}_k}$ in the $k$-th HARQ round reads as
\begin{align}\label{eqn:yk}
{{\bf y}_k} = \sqrt{P_k}{h_k}{{\bf x}_k} + {{\bf z}_k},
\end{align}
where the preceding definitions of $P_1$, $h_1$ and ${{\bf z}_1}$ also apply to $P_k$, $h_k$ and ${{\bf z}_k}$. At the receiver, the concatenated message ${\rm m}_{[k]}$ is decoded based on all the previous observations ${{\bf y}_1}, \cdots, {{\bf y}_k}$. The XP-HARQ process will be terminated only if the message ${\rm m}_{[k]}$ is reconstructed or the maximum number $K$ of HARQ rounds is reached.

\subsection{Outage Formulation}
The outage probability is recognized as the most pivotal performance metric of HARQ protocols. More precisely, the outage probability is defined as the probability of the event that the transmission rate vector lies outside the achievable region of XP-HARQ. For notational brevity, let $R_k$ define the incremental transmission rate introduced by ${\rm m}_k$. As proved in \cite{jabi2017adaptive}, the outage occurs if the accumulated mutual information is less than the accumulated rate. Precisely, the outage probability after $K$ HARQ rounds is formulated as
\begin{align}\label{eqn:p_out}
{P_{out,K}} = \Pr \left({I_1} < {R_1}, {I_2^\Sigma} < R_2^\Sigma, \cdots ,I_K^\Sigma  < R_K^\Sigma \right),
\end{align}
where $R_k^\Sigma  = \sum\nolimits_{l = 1}^k {{R_l}} $ and $I_k^\Sigma = \sum\nolimits_{l = 1}^k {I_l} $ denote the accumulated rate and the accumulated mutual information after $k$ HARQ rounds, respectively, $I_l =  {\log _2}(1 + {\gamma _l})$ represents the mutual information at the $l$-th transmission, and ${\gamma _l} = {{{{\left| {{h_l}} \right|}^2}{P_l}}}/{{{\sigma ^2}}}$ is the received signal-to-noise ratio (SNR) in the $l$-th HARQ round. It is worth noting that independent block fading channels and fixed length of sub-codewords across all HARQ rounds are assumed in the subsequent analysis for the analytical simplicity. The similar approach developed in this paper is also applicable to more complex scenarios, e.g., time-correlated fading channels and variable length XP-HARQ. 
\section{Outage Probability Analysis}\label{sec:out}
In what follows, the exact outage probability ${P_{out,K}}$ with $K=2$ is first derived in closed-form, which enables the asymptotic outage analysis with valuable insights. The similar results are then extended to a general case with $K>2$.
\subsection{The Case of $K=2$}
\subsubsection{Exact Outage Analysis}
According to \eqref{eqn:p_out}, the outage probability for $K=2$ is rewritten as
\begin{align}\label{eqn:outage_expk}
{P_{out,2}} 
 & ={{\text{E}}_{{{\gamma }_{2}}}}\left\{ \Pr \left( \left. {{\left| {{h}_{1}} \right|}^{2}}<{\gamma_{\rm th}{{\sigma }^{2}}}/{{{P}_{1}}} \right|{{\gamma }_{2}} \right) \right\} \notag\\
 & ={{\text{E}}_{{{\gamma }_{2}}}}\left\{ \left( 1-{{e}^{-\frac{\gamma_{\rm th}{{\sigma }^{2}}}{{{P}_{1}}}}} \right)u\left( {{2}^{{{R}_{1}}+{{R}_{2}}}}-1-{{\gamma }_{2}} \right) \right\},
\end{align}
where the second equality holds due to the exponential distribution of ${\left| {{h_1}} \right|^2}$, $\gamma_{\rm th} = {\min \left\{ {{2^{{R_1}}} - 1,{{{2^{{R_1} + {R_2}}}}}/{{(1 + {\gamma _2})}} - 1} \right\}}$, and $u( \cdot )$ stands for the unit step function. Since $h_k$ is Rayleigh distributed, the probability density function (PDF) of ${f_{{\gamma _2}}}(x) = {\sigma ^2}\exp ( - {\sigma ^2}x/{P_2})/{P_2}$. Accordingly, \eqref{eqn:outage_expk} can be expressed as
\begin{align}\label{eqn:outagek2_exp_expt}
{P_{out,2}}
 =&\left( {1 - {e^{ - \frac{{{\sigma ^2}}}{{{P_1}}}\left( {{2^{{R_1}}} - 1} \right)}}} \right)\left( {1 - {e^{ - \frac{{{\sigma ^2}}}{{{P_2}}}\left( {{2^{{R_2}}} - 1} \right)}}} \right) \notag\\
 &+ {e^{ - \frac{{{\sigma ^2}}}{{{P_2}}}\left( {{2^{{R_2}}} - 1} \right)}} - {e^{ - \frac{{{\sigma ^2}}}{{{P_2}}}\left( {{2^{{R_1} + {R_2}}} - 1} \right)}} \notag\\
 &-\underbrace {\frac{{{\sigma ^2}}}{{{P_2}}}{e^{\frac{{{\sigma ^2}}}{{{P_1}}} + \frac{{{\sigma ^2}}}{{{P_2}}}}}\int\nolimits_{{2^{{R_2}}}}^{{2^{{R_1} + {R_2}}}} {{e^{ - \frac{{{2^{{R_1} + {R_2}}}{\sigma ^2}}}{{z{P_1}}} - \frac{{{\sigma ^2}}}{{{P_2}}}z}}} dz}_{\triangleq \varphi \left( {{R_1},{R_2}} \right)}.
\end{align}

To derive $\varphi \left( {{R_1},{R_2}} \right)$ in closed-form, the integral of Mellin-Branes type of the exponential function ${e^{ - z}} = \frac{1}{{2\pi {\rm i}}}\int_{c - {\rm i}\infty }^{c + {\rm i}\infty } {\Gamma (s){z^{ - s}}ds}$ is invoked, where $c>0$, ${\left| {\arg z} \right| < \pi /2}$, and $\arg$ represents the argument of a complex number \cite[eq. 1.37]{mathai2009h}. 
After some algebraic manipulations, $\varphi \left( {{R_1},{R_2}} \right)$ is expressed as
\begin{align}\label{exp3}
&\varphi ({R_1},{R_2}) = 
{e^{\frac{{{\sigma ^2}}}{{{P_1}}} + \frac{{{\sigma ^2}}}{{{P_2}}}}}\frac{1}{{2\pi {\rm{i}}}}\int_{c - {\rm{i}}\infty }^{c + {\rm{i}}\infty } {\Gamma \left( s \right){{\left( {\frac{{{2^{{R_1} + {R_2}}}{\sigma ^4}}}{{{P_1}{P_2}}}} \right)}^{ - s}}} \notag\\
&\times \left( {\Gamma \left( {s + 1,\frac{{{\sigma ^2}}}{{{P_2}}}{2^{{R_2}}}} \right) - \Gamma \left( {s + 1,\frac{{{\sigma ^2}}}{{{P_2}}}{2^{{R_1} + {R_2}}}} \right)} \right)ds,
\end{align}
where $\Gamma (a)$, and $\Gamma (a,b)$ stands for the Gamma function and the upper incomplete Gamma function \cite[eq. 3.381.3]{gradshteyn1965table}, respectively. By identifying the integral of Mellin-Branes type in \eqref{exp3} with the generalized upper incomplete Fox's H function ${ H}_{p,q}^{m,n}\left[  \cdot  \right]$ \cite{yilmaz2009productshifted}, $\varphi ({R_1},{R_2})$ is obtained as
\begin{align}\label{eqn:varphi_ana_res}
&\varphi ({R_1},{R_2}) = {e^{\frac{{{\sigma ^2}}}{{{P_1}}} + \frac{{{\sigma ^2}}}{{{P_2}}}}}{{{ H}}}_{1,1}^{1,1}\left[ {\frac{{{2^{{R_1} + {R_2}}}{\sigma ^4}}}{{{P_1}{P_2}}}\left| {\begin{array}{*{20}{c}}
{\left( {0, - 1,\frac{{{\sigma ^2}}}{{{P_2}}}{2^{{R_2}}}} \right)}\\
{(0,1,0)}
\end{array}} \right.} \right] \notag\\
&-{e^{\frac{{{\sigma ^2}}}{{{P_1}}} + \frac{{{\sigma ^2}}}{{{P_2}}}}}{{{ H}}}_{1,1}^{1,1}\left[ {\frac{{{2^{{R_1} + {R_2}}}{\sigma ^4}}}{{{P_1}{P_2}}}\left| {\begin{array}{*{20}{c}}
{\left( {0, - 1,\frac{{{\sigma ^2}}}{{{P_2}}}{2^{{R_1} + {R_2}}}} \right)}\\
{(0,1,0)}
\end{array}} \right.} \right].
\end{align}
By substituting \eqref{eqn:varphi_ana_res} into \eqref{eqn:outagek2_exp_expt}, the outage probability ${P_{out,2}}$ can be gotten in closed-form. 
\subsubsection{Asymptotic Outage Analysis}
Due to the involvement of Fox's H function into the outage expression, it is notoriously difficult to reveal more insights about XP-HARQ. Therefore, we recourse to the asymptotic analysis of the outage probability in the high SNR regime, i.e., ${P_1/\sigma^2} , \cdots  , {P_K/\sigma^2}\to \infty$. To proceed, it is necessary to determine the asymptotic expression of $\varphi ({R_1},{R_2})$, as given by the following theorem.
\begin{theorem}\label{THEOREM I}
$\varphi ({R_1},{R_2})$ is asymptotically equal to
\begin{multline}\label{eqn:varphi_asy_fin}
\varphi \left( {{R_1},{R_2}} \right)  \simeq  {e^{\frac{{{\sigma ^2}}}{{{P_1}}}}}\left( {{e^{ - \frac{{{\sigma ^2}}}{{{P_2}}}\left( {{2^{{R_2}}} - 1} \right)}} - {e^{ - \frac{{{\sigma ^2}}}{{{P_2}}}\left( {{2^{{R_1} + {R_2}}} - 1} \right)}}} \right)\\
 - {e^{\frac{{{\sigma ^2}}}{{{P_1}}} + \frac{{{\sigma ^2}}}{{{P_2}}}}}\frac{{{\sigma ^4}}}{{{P_1}{P_2}}}{2^{{R_1} + {R_2}}}{R_1}\ln 2 + o\left( \frac{\sigma^4}{P_1P_2} \right),
\end{multline}
where $o(x)$ denotes the higher-order infinitesimal of $x$ as $x\to 0$.
\begin{proof}
Please see Appendix \ref{app_(10)}.
\end{proof}
\end{theorem}
Then, by substituting (\ref{eqn:varphi_asy_fin}) into (\ref{eqn:outagek2_exp_expt}) and using Taylor series of the exponential function $e^{x} =1 + x + o(x)$, keeping the highest-order terms leads to 
\begin{align}\label{eqn:pout_asy_k2}
&{{P}_{out,2}}  \simeq  \left( 1-{{e}^{-\frac{\left( {{2}^{{{R}_{1}}}}-1 \right){{\sigma }^{2}}}{{{P}_{1}}}}} \right)\left( 1-{{e}^{-\frac{{{\sigma }^{2}}}{{{P}_{2}}}\left( {{2}^{{{R}_{2}}}}-1 \right)}} \right)\notag\\
& +\left( {{e}^{-\frac{{{\sigma }^{2}}}{{{P}_{2}}}\left( {{2}^{{{R}_{2}}}}-1 \right)}}-{{e}^{-\frac{{{\sigma }^{2}}}{{{P}_{2}}}\left( {{2}^{{{R}_{1}}+{{R}_{2}}}}-1 \right)}} \right)\left( 1-{{e}^{\frac{{{\sigma }^{2}}}{{{P}_{1}}}}} \right)\notag\\
& +{{e}^{\frac{{{\sigma }^{2}}}{{{P}_{1}}}+\frac{{{\sigma }^{2}}}{{{P}_{2}}}}}\frac{{{2}^{{{R}_{1}}+{{R}_{2}}}}{{\sigma }^{4}}}{{{P}_{1}}{{P}_{2}}}{{R}_{1}}\ln 2+o\left(  \frac{\sigma^4}{P_1P_2} \right) \notag\\
& \simeq   \frac{\left( {{2}^{{{R}_{1}}+{{R}_{2}}}}{{R}_{1}}\ln2-\left( {{2}^{{{R}_{1}}}}-1 \right) \right){{\sigma }^{4}}}{{{P}_{1}}{{P}_{2}}}+o\left( \frac{\sigma^4}{P_1P_2} \right).
\end{align}

From \eqref{eqn:pout_asy_k2}, the diversity order of XP-HARQ can be readily obtained, where the diversity order is used to measure the degrees of freedom of communication systems. To be specific, the diversity order is mathematically given by
\begin{align}\label{div_ord}
{d_K} =  - \mathop {\lim }\limits_{{\bar \gamma} \to \infty } {{\log ({P_{out,K}})}}/{{\log ({\bar \gamma})}}.
\end{align}
where $\bar \gamma  \triangleq {P_1/\sigma^2} = \cdots  = {P_K/\sigma^2}$. By substituting (\ref{eqn:pout_asy_k2}) into (\ref{div_ord}), it is thus proved that the diversity order of XP-HARQ under $K=2$ is 2. Accordingly, full time diversity can be achieved by XP-HARQ.
\subsection{The Case of $K>2$}
Unfortunately, it is hardly impossible to derive the outage probability in a compact form for $K>2$. Therefore, we firstly resort to some bounds to approximate the outage probability. The asymptotic outage analysis is then performed to gain useful insights of XP-HARQ under general settings of $K$.
\subsubsection{Approximate Outage Analysis}
It is readily found from (\ref{eqn:p_out}) that the outage probability is bounded as
\begin{align}\label{eqn:out_bounds}
\underbrace {\Pr \left( {{I_1} < {R_1}, \cdots ,{I_K} < {R_K}} \right)}_{{P_{out,K,low}}} \le {P_{out,K}}
 \le \underbrace {\Pr \left( {I_K^\Sigma < {R_K^\Sigma}} \right)}_{{P_{out,K,up}}}.
\end{align}
With regard to the lower bound ${P_{out,K,low}}$, the independence among Rayleigh fading channels yields
\begin{equation}\label{eqn:out_lower_fin}
{P_{out,K,low}} = \prod\nolimits_{k = 1}^K {\left( {1 - {e^{ - \frac{{{\sigma ^2}}}{{{P_k}}}\left( {{2^{{R_k}}} - 1} \right)}}} \right)} ,
\end{equation}

Moreover, it can be seen that the upper bound ${P_{out,K,up}}$ is actually the outage probability of the conventional HARQ-IR, which is given by \eqref{eqn:out_upp}, as shown at the top of the next page,
\begin{figure*}
\begin{align}\label{eqn:out_upp}
{{P}_{out,K,up}}
& =Y_{K+1,1}^{1,K}\left[ \begin{matrix}
   (0,1,\frac{{{\sigma }^{2}}}{{{P}_{1}}},1),\cdots ,(0,1,\frac{{{\sigma }^{2}}}{{{P}_{K}}},1),(1,1,0,1)  \\
   (0,1,0,1)  \\
\end{matrix}\left| \frac{\prod\nolimits_{k=1}^{K}{{{{P}_{k}}}/{{{\sigma }^{2}}}}}{{{2}^{R_{K}^{\Sigma }}}} \right. \right].
\end{align}
\vspace*{-4pt}
\end{figure*}
where $Y_{p,q}^{m,n}\left[  \cdot  \right]$ is the generalized Fox's H function \cite{shi2017asymptotic}.

\subsubsection{Asymptotic Outage Analysis}
To extract more insights, the asymptotic outage analysis for $K>2$ is carried out in the following. To this end, ${P_{out,K}}$ is rewritten by using \eqref{eqn:p_out} as
\begin{align}\label{eqn:out_gene_1}
{P_{out,K}} 
 &= \Pr \left( {1 + {\gamma _1} < {2^{{R_1}}}, \cdots ,\prod\nolimits_{k = 1}^K {\left( {1 + {\gamma _k}} \right)}  < {2^{R_K^\Sigma }}} \right),
\end{align}
By make change of variable ${x_k} = \prod\nolimits_{l = 1}^k {\left( {1 + {\gamma _l}} \right)}  $, 
(\ref{eqn:out_gene_1}) can be expressed as
\begin{multline}\label{eqn:out_toxk}
{P_{out,K}} = \Pr \left( {x_0 < {x_1} < {2^{{R_1}}}, \cdots ,{x_{K - 1}} < {x_K} < {2^{R_K^\Sigma }}} \right) \\
 = \int\nolimits_{x_0}^{{2^{{R_1}}}} { \cdots {\int\nolimits_{{x_{K - 1}}}^{{2^{R_K^\Sigma }}} {{f_{{x_1}, \cdots ,{x_K}}}\left( {{x_1}, \cdots ,{x_K}} \right)d{x_1} \cdots d{x_K}} } },
\end{multline}
where $x_0=1$. Therefore, we have to determine the joint PDF of ${\bf x} = (x_1, \cdots, {{x}_{K}})$. By capitalizing on Jacobian transform, the joint PDF of $\bf x$ can be obtained as ${{f}_{\bf x}}({{x}_{1}},\cdots ,{{x}_{K}})={{f}_{{\bs \gamma}}}({{\gamma }_{1}},\cdots ,{{\gamma }_{K}})\det({\bf J})$, where the joint PDF of ${\bs \gamma}=({{\gamma }_{1}},\cdots,{{\gamma }_{K}})$ is ${f_{{\bs \gamma}}}\left( {{\gamma _1}, \cdots ,{\gamma _K}} \right) =  \prod\nolimits_{k = 1}^K {{\sigma ^2}\exp \left( { - {\sigma ^2}{\gamma _k}/{P_k}} \right)/{P_k}} $ due to the independence among channel coefficients, and the matrix $\bf J$ is given by 
\begin{align}\label{eqn:jacobian}
{\bf J} 
 &=  \left( {\begin{array}{*{20}{c}}
1&0& \cdots &0\\
{ - {{{x_2}}}{{{x_1}^{-2}}}}&{{{{x_1}}}^{-1}}& \cdots &0\\
0& \vdots & \ddots & \vdots \\
0&0& \cdots &{{{{x_{K - 1}}}}^{-1}}
\end{array}} \right).
\end{align}
%

Accordingly, we get ${f_{\bf x}}({x_1}, \cdots ,{x_K}) = \prod\nolimits_{i = 1}^{K - 1} {\frac{1}{{{x_i}}}} \prod\nolimits_{k = 1}^K {\frac{{{\sigma ^2}}}{{{P_k}}}} \exp \left( { - \frac{{{\sigma ^2}}}{{{P_k}}}\left( {\frac{{{x_k}}}{{{x_{k - 1}}}} - 1} \right)} \right)$.  Putting the joint PDF of $\bf x$ into \eqref{eqn:out_toxk} gives rise to (\ref{eqn:out_toxk_rew1}), as shown at the top of the following page.
\begin{figure*}
\begin{align}\label{eqn:out_toxk_rew1}
{P_{out,K}} &= \prod\limits_{k = 1}^K {\frac{{{\sigma ^2}}}{{{P_k}}}{e^{\frac{{{\sigma ^2}}}{{{P_k}}}}}} \int\nolimits_1^{{2^{{R_1}}}} { \cdots \int\nolimits_{{x_{K - 2}}}^{{2^{R_{K - 1}^\Sigma }}} {\int\nolimits_{{x_{K - 1}}}^{{2^{R_K^\Sigma }}} {{e^{ - \frac{{{\sigma ^2}}}{{{P_k}}}\frac{{{x_K}}}{{{x_{K - 1}}}}}}\prod\limits_{k = 1}^{K - 1} {\frac{1}{{{x_k}}}{e^{ - \frac{{{\sigma ^2}}}{{{P_k}}}\frac{{{x_k}}}{{{x_{k - 1}}}}}}} d{x_1} \cdots d{x_{K - 1}}d{x_K}} } } ,
\end{align}
\hrulefill
\end{figure*}
By using Taylor expansion of the exponential functions together with some trivial manipulations, it follows that
\begin{align}\label{eqn:out_gen_mani}
&{P_{out,K}} = \prod\limits_{k = 1}^K {\frac{{{\sigma ^2}}}{{{P_k}}}{e^{\frac{{{\sigma ^2}}}{{{P_k}}}}}} \sum\limits_{{n_1}, \cdots ,{{\rm{n}}_K}{\rm{ = }}0}^\infty  {\prod\limits_{k = 1}^K {{{{\left( { - \frac{{{\sigma ^2}}}{{{P_k}}}} \right)}^{{n_k}}}}\frac{1}{{{n_k}!}}} }  \notag\\
& \times \int\nolimits_{{x_0}}^{{2^{{R_1}}}} { \cdots \int\nolimits_{{x_{K - 2}}}^{{2^{R_{K - 1}^\Sigma }}} {\prod\limits_{k = 1}^{K - 1} {{x_k}^{{n_k} - {n_{k + 1}} - 1}} } } d{x_1} \cdots d{x_{K - 1}}\notag\\
& \times {{\left( {{2^{\left( {{n_K} + 1} \right)R_K^\Sigma }} - {x_{K - 1}}^{{n_K} + 1}} \right)}}/{({{n_K} + 1})}.
\end{align}

Based on the representation of the summation of the infinite terms in \eqref{eqn:out_gen_mani}, the asymptotic outage expression under high SNR (i.e., ${P_1/\sigma^2} , \cdots  , {P_K/\sigma^2}\to \infty$) can be obtained by ignoring the higher order terms of $\prod\nolimits_{k=1}^{K}{\exp{(-{\sigma^2/P_k})}}$, that is, the dominant term is the one with ${{n}_{1}},\cdots ,{{n}_{K}}=0$. Thus, it gives 
%
\begin{align}\label{eqn:out_asy_exa}
&{P_{out,K}} \simeq  \notag\\
&\prod\limits_{k = 1}^K {\frac{{{\sigma ^2}}}{{{P_k}}}} \underbrace {\int\limits_{{x_0}}^{{2^{{R_1}}}}  \cdots  \int\limits_{{x_{K - 2}}}^{{2^{R_{K - 1}^\Sigma }}} {\prod\limits_{k = 1}^{K - 1} {{x_k}^{ - 1}} } \left( {{2^{R_K^\Sigma }} - {x_{K - 1}}} \right)d{x_1} \cdots d{x_{K - 1}}}_{{\hbar _{K,1}}\left( {{x_0}} \right)},
\end{align}
where ${\hbar _{K,1}}\left( x_0 \right)$ can be recursively computed with the following theorem.
\begin{theorem}\label{THEOREM_II}
${\hbar _{K,k}}\left( {{x_{k - 1}}} \right)$ satisfies the following recursive relationship as
\begin{multline}\label{eqn:out_h_def}
{\hbar _{K,k}}\left( {{x_{k - 1}}} \right) = \int\nolimits_{{x_{k - 1}}}^{{2^{R_k^\Sigma }}} {{x_k}^{ - 1}{\hbar _{K,k + 1}}\left( {{x_k}} \right)d{x_k}} \\
 = {\left( { - 1} \right)^{K - k + 1}}{x_{k - 1}} + \sum\nolimits_{{\mathop{ i}\nolimits}  = 0}^{K - k} {{c_{k,{\mathop{i}\nolimits} }}{{\left( {\ln \left( {{x_{k - 1}}} \right)} \right)}^{\mathop{ i}\nolimits} }} ,
\end{multline}
where ${c_{k,i}} =  - {c_{k + 1,i - 1}}/i,i \in [1,K - k]$, ${c_{k,0}} = \sum\nolimits_{i = 0}^{K - k - 1} {{{{c_{k + 1,i}}{{(\ln ({2^{R_k^\Sigma }}))}^{i + 1}}} \mathord{\left/{\vphantom {{{c_{k + 1,i}}{{(\ln({2^{R_k^\Sigma }}))}^{i + 1}}} {(i + 1)}}} \right.\kern-\nulldelimiterspace} {(i + 1)}}}  + {( - 1)^{K - k}}{2^{R_k^\Sigma }}$, ${c_{K,0}} = {2^{R_K^\Sigma }}$, and ${\hbar _{K,K}}\left( {{x_{K - 1}}} \right) = {2^{R_K^\Sigma }} - {x_{K - 1}}$.
\begin{proof}
Please see Appendix \ref{app_23}.
\end{proof}
\end{theorem}

Besides, by substituting (\ref{eqn:out_asy_exa}) into (\ref{div_ord}), the diversity order of XP-HARQ under $K>2$ is $d_K=K$. This indicates that the full time diversity can be attained by XP-HARQ.
\section{NUMERICAL RESULTS}\label{NUM_RES}
In this section, numerical results are provided for verifications. Unless otherwise indicated, we assume equal power allocation for XP-HARQ (i.e., $P_1=\cdots=P_K=P$), which results in the same average received SNR (i.e., ${P_1/\sigma^2} , \cdots  , {P_K/\sigma^2}\triangleq \bar \gamma$). For notional convenience, the transmission rates are defined by a vector ${\bf R}=(R_1,\cdots,R_K)$.

\begin{figure}[htbp]
\centering
\subfigure[$K=2$]
{
\includegraphics[width=3.7cm]{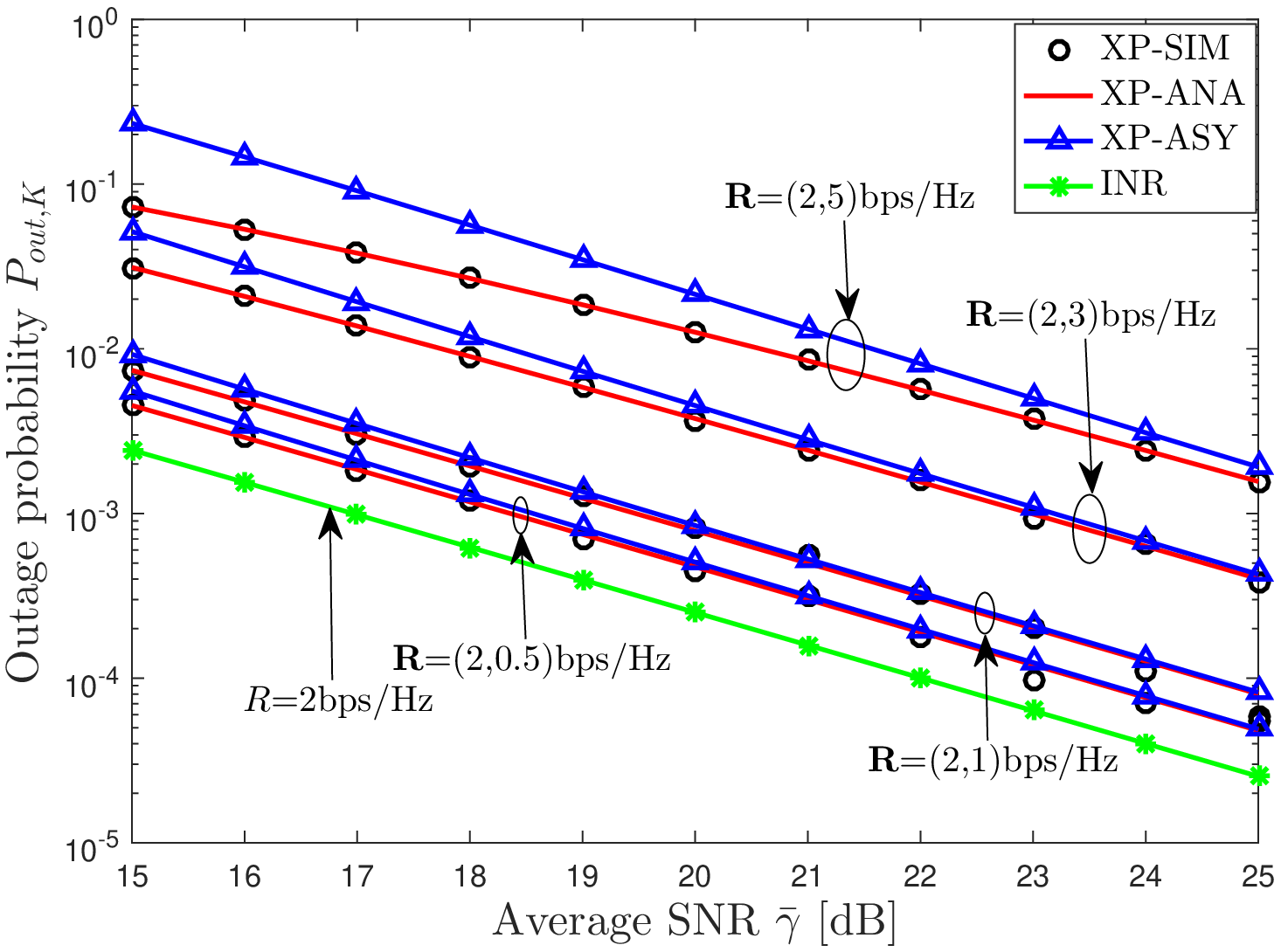}
\label{fig_1_a}
}
\quad
\subfigure[$K=3$]
{
\includegraphics[width=3.7cm]{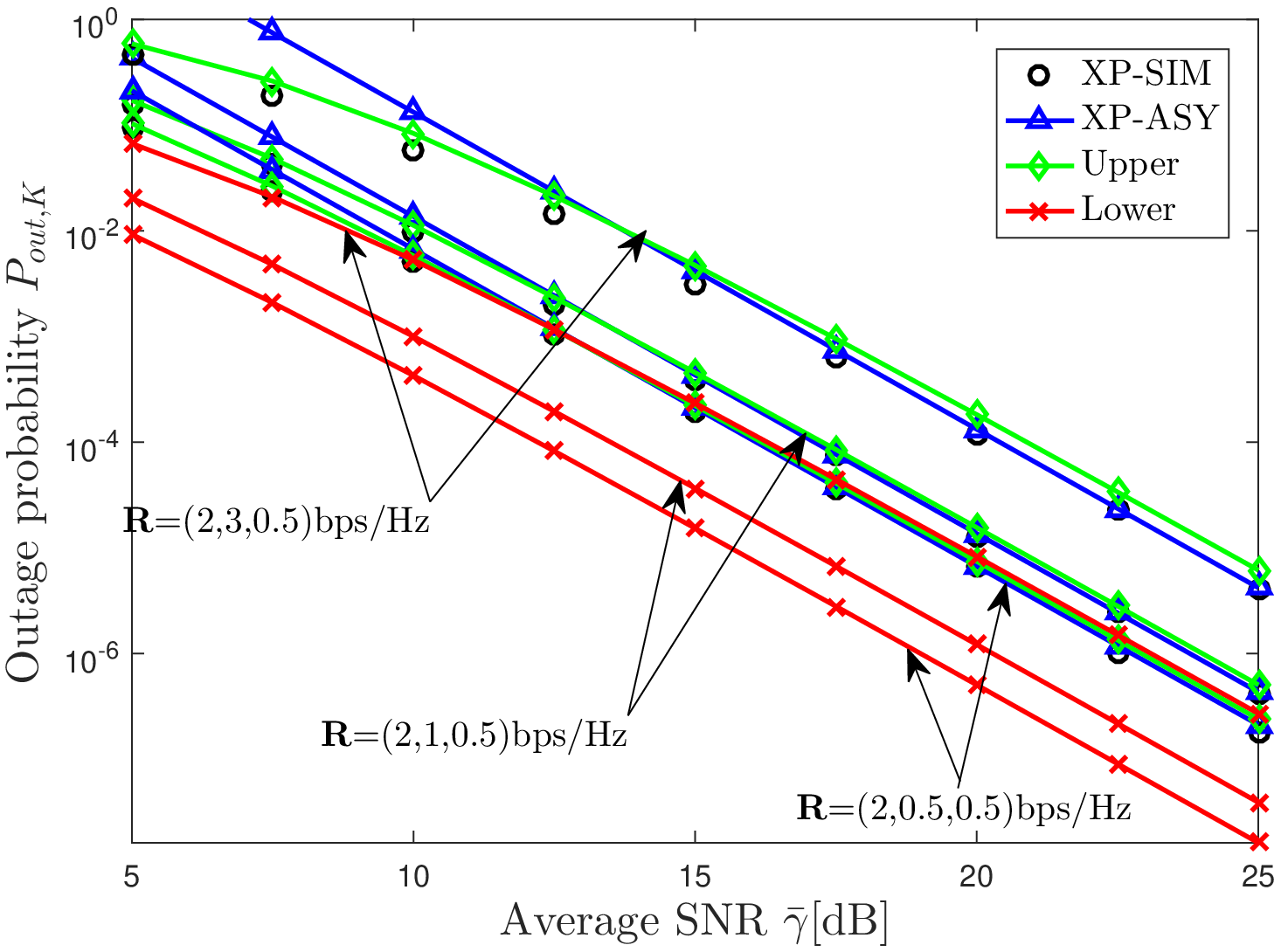}
\label{fig_1_b}
}
\caption{The outage probability versus the average received SNR $\bar \gamma$}
\label{fig:1}
\end{figure}

Fig. \ref{fig_1_a} depicts the relationship between the outage probability ${P_{out,2}}$ and the average SNR for different $\bf R$, where the labels ``SIM'', ``ANA'' and ``ASY'' represent the simulated, the exact and the asymptotic results, respectively, and ``INR'' refers to INR-HARQ. It can be seen that the exact and simulated results are in perfect agreement. Besides, the exact results approach to the asymptotic ones under high SNR. The observations confirm the correctness of our analysis. Furthermore, the outage probability declines with the average SNR. In fact, the declining rate of the outage curves in the log-log scale is equal to the diversity order. With this relationship, the analysis of the diversity order can be proved. In addition, it can be observed that the outage curves under high SNR are parallel to each other even if the values of $\bf R$ are different. 
Furthermore, Fig. \ref{fig_1_b} presents the outage probability versus the average SNR for $K=3$, where the labels ``Upper'' and ``Lower'' represent the upper and the lower bounds of outage probability, respectively. 
Besides, it is easily found that the simulated results lies in between the lower and upper bounds of the outage probability, and the simulated results tend to the asymptotic ones under high SNR. These observations corroborate our analytical results. Similarly to Fig. \ref{fig_1_a}, it can be found that the decreasing rate of the outage curves is $d_K=3$ regardless of the values of $\bf R$. Moreover, it can be observed from Fig. \ref{fig_1_b} that the gap between the upper bounds and the simulated results is negligible for small $R_2$ and $R_3$.

\begin{figure}[htbp]
\centering
\begin{minipage}[t]{0.47\linewidth}
\includegraphics[width=3.7cm]{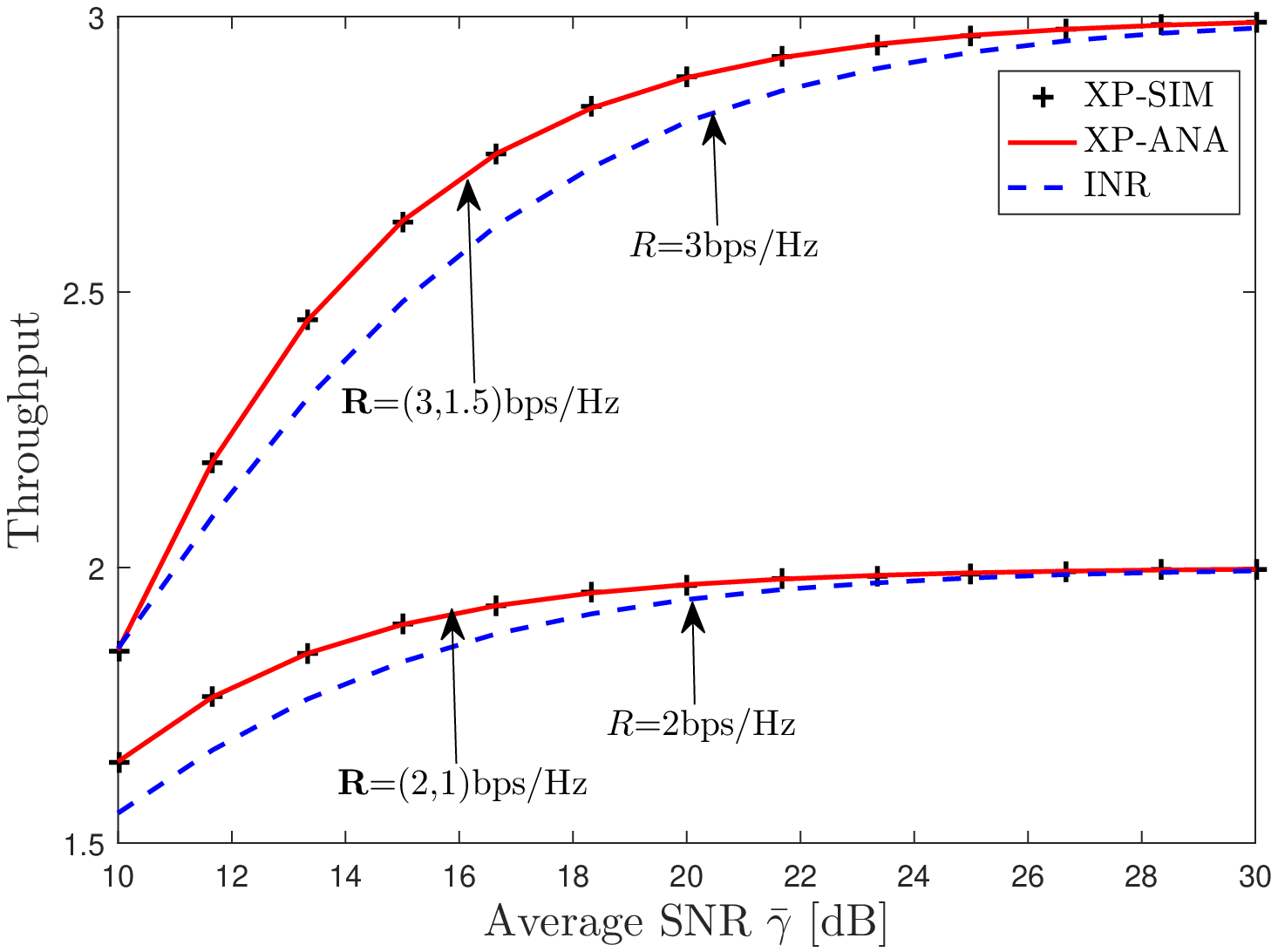}
\caption{The throughput versus the average SNR $\bar \gamma$ for $K=2$.}\label{fig_2_a}
\end{minipage}
\quad
\begin{minipage}[t]{0.47\linewidth}
\includegraphics[width=3.7cm]{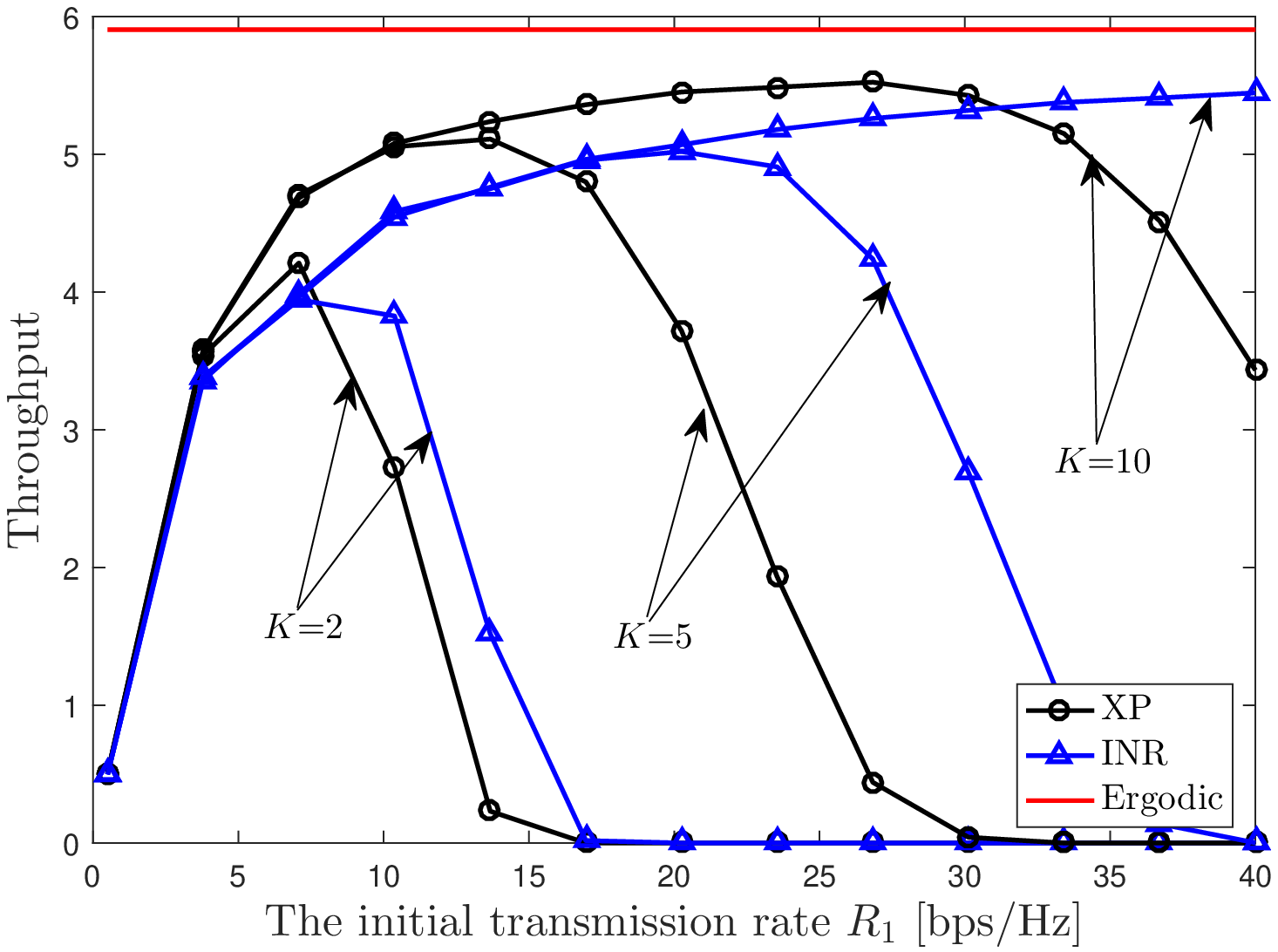}
\caption{The throughput versus the initial transmission rate $R_1$ for $R_2=\cdots=R_K=2$~bps/Hz and $\bar \gamma=20$~dB.}
\label{fig_2_b}
\end{minipage}
\end{figure}

Fig. \ref{fig_2_a} investigates the throughout of XP-HARQ and INR-HARQ against the average SNR, where the explicit expressions of the throughput of HARQ schemes are given by \cite{jabi2017adaptive}. The analytical results coincide with the simulated results. It is also observed that XP-HARQ is superior to INR-HARQ in terms of the throughput.

In Fig. \ref{fig_2_b}, the throughput and the ergodic capacity of XP-HARQ and INR-HARQ are plotted against the initial transmission rate. As $K$ and $R$ increase, both the maximum throughputs of XP-HARQ and INR-HARQ tend to the ergodic capacity. This is consistent with the results in \cite{Makki2010}. Besides, XP-HARQ outperforms INR-HARQ in terms of the maximum throughput. 

\section{Conclusions}\label{sec:con}
To theoretically evaluate the outage performance of XP-HARQ, we have first derived the exact outage probability in the case of $K=2$. Unfortunately, due to difficulty of dealing with the multi-fold integral, the similar approach has not been applied to get a general expression for the outage probability of XP-HARQ with $K>2$. Besides, the asymptotic outage analysis has revealed that the full time diversity is attainable by XP-HARQ. Moreover, the simple form of the asymptotic outage expression facilitates the optimal design of XP-HARQ system, which will be discussed in our future works. Finally, the numerical results have demonstrated the correctness of our analysis.
\appendices
\section{Proof of Theorem \ref{THEOREM I}}\label{app_(10)}
By using the residue theorem, \eqref{eqn:varphi_ana_res} can be expanded as
\begin{align}\label{eqn:res_the}
&\varphi \left( {{R_1},{R_2}} \right) = {e^{\frac{{{\sigma ^2}}}{{{P_1}}} + \frac{{{\sigma ^2}}}{{{P_2}}}}}\sum\nolimits_{j =  - \infty }^0 {} \notag\\
&\left( {\begin{array}{*{20}{l}}
{{\rm{Res}}\left\{ {\Gamma (s)\Gamma \left( {s + 1,\frac{{{\sigma ^2}}}{{{P_2}}}{2^{{R_2}}}} \right){{\left( {\frac{{{2^{{R_1} + {R_2}}}{\sigma ^4}}}{{{P_1}{P_2}}}} \right)}^{ - s}},j} \right\} - }\\
{{\rm{Res}}\left\{ {\Gamma (s)\Gamma \left( {s + 1,\frac{{{\sigma ^2}}}{{{P_2}}}{2^{{R_1} + {R_2}}}} \right){{\left( {\frac{{{2^{{R_1} + {R_2}}}{\sigma ^4}}}{{{P_1}{P_2}}}} \right)}^{ - s}},j} \right\}}
\end{array}} \right),
\end{align}
where ${\mathop{\rm Res}\nolimits} (f(s),{s=s_j})$ denotes the residue of $f(s)$ at pole ${s_j}$. By applying dominant term approximation ({i.e., $P_1/\sigma^2, P_2/\sigma^2\to \infty$), we only need to consider the poles at ${s} = 0$ and ${s} = -1$ in \eqref{eqn:res_the}. Moreover, by considering that $\Gamma(s)$ is not well defined at $s=0 $ and $ -1$, the identities $s\Gamma(s)=\Gamma(s+1)$ and $(s+1)\Gamma(s)=\Gamma(s+2)/s$ are applied to simplify \eqref{eqn:res_the}. Accordingly, one has

\begin{align}\label{eqn:varphi_asy_res}
& \varphi ({{R}_{1}},{{R}_{2}}) \simeq {e^{\frac{{{\sigma ^2}}}{{{P_1}}} + \frac{{{\sigma ^2}}}{{{P_2}}}}} \times \notag\\
&\left( {\begin{array}{*{20}{l}}
{\Gamma \left( {1,\frac{{{\sigma ^2}}}{{{P_2}}}{2^{{R_2}}}} \right) - \Gamma \left( {0,\frac{{{\sigma ^2}}}{{{P_2}}}{2^{{R_2}}}} \right)\frac{{{2^{{R_1} + {R_2}}}{\sigma ^4}}}{{{P_1}{P_2}}} - }\\
{\Gamma \left( {1,\frac{{{\sigma ^2}}}{{{P_2}}}{2^{{R_1} + {R_2}}}} \right) + \Gamma \left( {0,\frac{{{\sigma ^2}}}{{{P_2}}}{2^{{R_1} + {R_2}}}} \right)\frac{{{2^{{R_1} + {R_2}}}{\sigma ^4}}}{{{P_1}{P_2}}}}
\end{array}} \right).
\end{align}
It is noticed from \cite[eq. 8.352.7]{gradshteyn1965table} that $\Gamma \left( {1, x } \right)$ is an exponential function $e^{-x}$. Moreover, it is found from \cite[eq. 8.359.1]{gradshteyn1965table} that $\Gamma \left( {0, x } \right)$ reduces to an exponential integral function $-{\rm Ei}(-x)$. Then using the asymptotic expansion \cite[eq. 8.214.1]{gradshteyn1965table} as $x\to 0$ leads to $\Gamma \left( {0, x } \right) \simeq -\ln(x)$. Hereby, (\ref{eqn:varphi_asy_res}) can be derived as \eqref{eqn:varphi_asy_fin}.

\section{Proof of Theorem \ref{THEOREM_II}}\label{app_23}
Theorem \ref{THEOREM_II} can be proved by induction. Clearly, \eqref{eqn:out_h_def} holds for $k=K$, because ${\hbar _{K,K}}\left( {{x_{K - 1}}} \right) = {2^{R_K^\Sigma }} - {x_{K - 1}}$. To prove \eqref{eqn:out_h_def}, we assume that the following identity holds
\begin{equation}\label{eqn:def_h_conj}
{\hbar _{K,k + 1}}\left( {{x_k}} \right) = {\left( { - 1} \right)^{K - k}}{x_k} + \sum\nolimits_{i = 0}^{K - k - 1} {{c_{k + 1,i}}{{\left( {\ln {{x_k}} } \right)}^ i}}
\end{equation}
Substituting (\ref{eqn:def_h_conj}) into the definition of ${\hbar _{K,k}}\left( {{x_{k - 1}}} \right)$ in (\ref{eqn:out_h_def})  yields
\begin{align}\label{eqn:h_K_k_x}
&{\hbar _{K,k}}\left( {{x_{k - 1}}} \right) = \int\nolimits_{{x_{k - 1}}}^{{2^{R_k^\Sigma }}} {{x_k}^{ - 1}{\hbar _{K,k + 1}}\left( {{x_k}} \right)d{x_k}} \notag\\
 &=  {\left( { - 1} \right)^{K - k + 1}}{x_{k - 1}} + \sum\nolimits_{i = 0}^{K - k - 1} {{c_{k + 1,i}}\frac{{{{\left( {\ln {2^{R_k^\Sigma }}} \right)}^{i + 1}}}}{{i + 1}}} \notag\\
 &+ {\left( { - 1} \right)^{K - k}}{2^{R_k^\Sigma }} - \sum\nolimits_{i = 1}^{K - k} {\frac{{{c_{k + 1,i - 1}}}}{{i + 1}}{{\left( {\ln {x_{k - 1}}} \right)}^i}}.
\end{align}
By comparing \eqref{eqn:h_K_k_x} with (\ref{eqn:out_h_def}), the recursive relationship between the coefficients $c_{k,{i}}$ can be obtained.

\bibliographystyle{ieeetran}
\bibliography{manuscript_1}

\begin{thebibliography}{10}
\providecommand{\url}[1]{#1}
\csname url@samestyle\endcsname
\providecommand{\newblock}{\relax}
\providecommand{\bibinfo}[2]{#2}
\providecommand{\BIBentrySTDinterwordspacing}{\spaceskip=0pt\relax}
\providecommand{\BIBentryALTinterwordstretchfactor}{4}
\providecommand{\BIBentryALTinterwordspacing}{\spaceskip=\fontdimen2\font plus
\BIBentryALTinterwordstretchfactor\fontdimen3\font minus
  \fontdimen4\font\relax}
\providecommand{\BIBforeignlanguage}[2]{{%
\expandafter\ifx\csname l@#1\endcsname\relax
\typeout{** WARNING: IEEEtran.bst: No hyphenation pattern has been}%
\typeout{** loaded for the language `#1'. Using the pattern for}%
\typeout{** the default language instead.}%
\else
\language=\csname l@#1\endcsname
\fi
#2}}
\providecommand{\BIBdecl}{\relax}
\BIBdecl

\bibitem{8636206}
G.~J. Sutton, J.~Zeng, R.~P. Liu, W.~Ni, D.~N. Nguyen, B.~A. Jayawickrama,
  X.~Huang, M.~Abolhasan, Z.~Zhang, E.~Dutkiewicz, and T.~Lv, ``Enabling
  technologies for ultra-reliable and low latency communications: From {PHY}
  and {MAC} layer perspectives,'' \emph{IEEE Commun. Surv. Tutor.}, vol.~21,
  no.~3, pp. 2488--2524, 3rd Quart., 2019.

\bibitem{9662195}
H.~Ding and K.~G. Shin, ``Context-aware beam tracking for {5G} {mmWave} {V2I}
  communications,'' \emph{IEEE Trans. Mobile Comput.}, vol.~PP, no.~99, pp.
  1--1, 2021.

\bibitem{ahmed2021hybrid}
A.~Ahmed, A.~Al-Dweik, Y.~Iraqi, H.~Mukhtar, M.~Naeem, and E.~Hossain, ``Hybrid
  automatic repeat request {(HARQ)} in wireless communications systems and
  standards: A contemporary survey,'' \emph{IEEE Commun. Surv. Tutor.},
  vol.~23, no.~4, pp. 2711--2752, Jul. 2021.

\bibitem{jabi2015multipacket}
M.~Jabi, A.~El~Hamss, L.~Szczecinski, and P.~Piantanida, ``Multipacket hybrid
  {ARQ}: Closing gap to the ergodic capacity,'' \emph{IEEE Trans. Commun.},
  vol.~63, no.~12, pp. 5191--5205, Oct. 2015.

\bibitem{mheich2020performance}
Z.~Mheich, W.~Yu, P.~Xiao, A.~U. Quddus, and A.~Maaref, ``On the performance of
  {HARQ} protocols with blanking in {NOMA} systems,'' \emph{IEEE Trans.
  Wireless Commun.}, vol.~19, no.~11, pp. 7423--7438, Jul. 2020.

\bibitem{khreis2020multi}
{{A. Khreis, F. Bassi, P. Ciblat and P. Duhamel}}, ``Multi-layer {HARQ} with
  delayed feedback,'' \emph{IEEE Trans. Wireless Commun.}, vol.~19, no.~9, pp.
  6224--6237, Sep. 2020.

\bibitem{nadeem2021non}
F.~Nadeem, M.~Shirvanimoghaddam, Y.~Li, and B.~Vucetic, ``Non-orthogonal {HARQ}
  for {URLLC}: Design and analysis,'' \emph{IEEE Internet Things J.}, Mar.
  2021.

\bibitem{4202098}
C.~Hausl and A.~Chindapol, ``Hybrid {ARQ} with cross-packet channel coding,''
  \emph{IEEE Commun. Lett.}, vol.~11, no.~5, pp. 434--436, May 2007.

\bibitem{jabi2017boost}
M.~Jabi, E.~Pierre-Doray, L.~Szczecinski, and M.~Benjillali, ``How to boost the
  throughput of {HARQ} with off-the-shelf codes,'' \emph{IEEE Trans. Commun.},
  vol.~65, no.~6, pp. 2319--2331, Jun. 2017.

\bibitem{jabi2018amc}
M.~Jabi, L.~Szczecinski, M.~Benjillali, A.~Benyouss, and B.~Pelletier, ``{AMC}
  and {HARQ}: How to increase the throughput,'' \emph{IEEE Trans. Commun.},
  vol.~66, no.~7, pp. 3136--3150, Feb. 2018.

\bibitem{liang2018efficient}
H.~Liang, A.~Liu, Y.~Zhang, and X.~Liang, ``Efficient design of multi-packet
  hybrid {ARQ} transmission scheme based on polar codes,'' \emph{IEEE Access},
  vol.~6, pp. 31\,564--31\,570, Jun. 2018.

\bibitem{jabi2017adaptive}
M.~Jabi, A.~Benyouss, M.~Le~Treust, E.~Pierre-Doray, and L.~Szczecinski,
  ``Adaptive cross-packet {HARQ},'' \emph{IEEE Trans. Commun.}, vol.~65, no.~5,
  pp. 2022--2035, May 2017.

\bibitem{mathai2009h}
A.~Mathai, R.~K. Saxena, and H.~J. Haubold, \emph{The H-function}.\hskip 1em
  plus 0.5em minus 0.4em\relax Springer, 2009.

\bibitem{gradshteyn1965table}
I.~S. Gradshteyn, I.~M. Ryzhik, A.~Jeffrey, D.~Zwillinger, and S.~Technica,
  \emph{Table of integrals, series, and products}, 7th~ed.\hskip 1em plus 0.5em
  minus 0.4em\relax Academic press New York, 2007.

\bibitem{yilmaz2009productshifted}
F.~Yilmaz and M.-S. Alouini, ``Product of shifted exponential variates and
  outage capacity of multicarrier systems,'' in \emph{Proc. European Wireless
  Conference (EW'09)}, May 2009, pp. 282--286.

\bibitem{shi2017asymptotic}
Z.~Shi, S.~Ma, G.~Yang, K.-W. Tam, and M.~Xia, ``Asymptotic outage analysis of
  {HARQ-IR} over time-correlated nakagami-$ m $ fading channels,'' \emph{IEEE
  Trans. Wireless Commun.}, vol.~16, no.~9, pp. 6119--6134, Sep. 2017.

\bibitem{Makki2010}
B.~Makki and T.~Eriksson, ``On the average rate of quasi-static fading channels
  with {ARQ} and {CSI} feedback,'' \emph{IEEE Commun. Lett.}, vol.~14, no.~9,
  pp. 806--808, Sep. 2010.

\end{thebibliography}

\end{document}